\documentclass{article}
\usepackage{spconf,amsmath,graphicx}

\usepackage[utf8]{inputenc}
\usepackage{graphicx}
\usepackage{amsmath, amssymb}
\usepackage[printonlyused,nolist]{acronym}
\usepackage{xcolor}
\usepackage{algorithm}
\usepackage{algpseudocode}
\usepackage{etoolbox}
\usepackage{multirow}
\usepackage{tikz}
\usepackage{subcaption}
\usepackage{url}
\usepackage{graphicx}
\usepackage{tikz}
\usepackage{pgfplots}
\usepackage{optidef}
\usepackage{multicol}

\title{Noise-Robust Adaptation Control for Supervised Acoustic System Identification Exploiting A Noise Dictionary}
\name{Thomas Haubner, Andreas Brendel, Mohamed Elminshawi, and Walter Kellermann\thanks{This work was partially funded by the DFG -- 282835863 -- within the Research Unit FOR2457 "Acoustic Sensor Networks".}}

\address{Multimedia Communications and Signal Processing,\\Friedrich-Alexander University Erlangen-N\"urnberg,\\Cauerstr. 7, 91058 Erlangen, Germany,\\thomas.haubner@fau.de}

\DeclareMathOperator*{\w}{\boldsymbol{w}}
\DeclareMathOperator*{\X}{\boldsymbol{X}}
\DeclareMathOperator*{\bP}{\boldsymbol{P}}
\DeclareMathOperator*{\e}{\boldsymbol{e}}
\DeclareMathOperator*{\C}{\boldsymbol{C}}
\DeclareMathOperator*{\y}{\boldsymbol{y}}
\DeclareMathOperator*{\I}{\boldsymbol{I}}
\DeclareMathOperator*{\bs}{\boldsymbol{s}}
\DeclareMathOperator*{\herm}{\text{H}}

\DeclareMathOperator*{\dicMat}{\boldsymbol{T}}
\DeclareMathOperator*{\actVec}{\boldsymbol{v}}

\begin{acronym}
	\acro{OSASI}{Online Supervised Acoustic System Identification}
	\acro{SNR}{Signal-to-Noise Ratio}
	\acro{SSFDAF}{State-Space Frequency-Domain Adaptive Filter}
	\acro{MCSSFDAF}{Multichannel State-Space Frequency-Domain Adaptive Filter}
	\acro{SSFDAF-NMF}{State-Space Frequency-Domain Adaptive Filter with an Nonnegative Matrix Factorization Noise Model}
	\acro{NMF}{NMF}
	\acro{IS}{Itakura-Saito}
	\acro{ML}{Maximum-Likelihood}
	\acro{EM}{Expectation-Maximization}
	\acro{AEC}{acoustic echo cancellation}
	\acro{PF}{Post Filter}
	\acro{LMS}{Least Mean Square}
	\acro{NLMS}{Normalized Least Mean Square}
	\acro{RLS}{Recursive Least Squares}
	\acro{DFT}{Discrete Fourier Transform}
	\acro{DTFT}{Discrete Time Fourier Transform}
	\acro{MAP}{Maximum A Posteriori}
	\acro{MMSE}{Minimum Mean Square Error}
	\acro{STFT}{Short-Time Fourier Transform}
	\acro{PSD}{Power Spectral Density}
	\acro{ERLE}{Echo Return Loss Enhancement}
	\acro{RIR}{room impulse response}
	\acro{FIR}{Finite Impulse Response}
	\acro{IR}{Impulse Response}
	\acro{ASR}{Automatic Speech Recognition}
	\acro{SS}{Source Separation}
	\acro{SC}{Single-Channel}
	\acro{MC}{Multichannel}
	\acro{IS-NMF}{Itakura-Saito NMF}
	\acro{pdf}{probability density function}
	\acro{VSSS}{Variable Step Size Selection}
	\acro{VSS}{Variable Step Size}
	\acro{FD}{frequency-domain}
	\acro{TD}{time-domain}
	\acro{MM}{Minorize-Maximization}
\end{acronym}

\algdef{SE}[SUBALG]{SubAlgorithm}{EndSubAlgorithm}{}{}%

\newtoggle{long}

\newcommand{\commentTHa}[1]{\textcolor{black}{#1}}
\newcommand{\commentTHc}[1]{\textcolor{black}{#1}}
\newcommand{\commentTHd}[1]{\textcolor{black}{#1}}
\newcommand{\commentTHf}[1]{\textcolor{black}{#1}}

\toggletrue{long}

\begin{document}
	\ninept
	\maketitle
	
	\begin{abstract}
		We present a noise-robust adaptation control strategy for block-online supervised acoustic system identification by exploiting a noise dictionary. The proposed algorithm takes advantage of the pronounced spectral structure which characterizes many types of interfering noise signals. We model the noisy observations by a linear Gaussian \acl{DFT}-domain state space model whose parameters are estimated by an online generalized \acl{EM} algorithm. Unlike all other state-of-the-art approaches we suggest to model the covariance matrix of the observation \acl{pdf} by a dictionary model. 
		We propose to learn the noise dictionary from training data, which can be gathered either offline or online whenever the system is not excited, while we infer the activations continuously. \commentTHc{The proposed algorithm represents a novel machine-learning-based approach to noise-robust adaptation control which allows for faster convergence in applications characterized by high-level and non-stationary interfering noise signals and abrupt system changes.}
	\end{abstract}
	
	%
	%
	%
	\section{Introduction}
	\label{sec:intro}
	\ac{OSASI} is required for many modern hands-free acoustic human-machine interfaces, e.g., for the purpose of echo cancellation \cite{enzner_acoustic_2014}.
	%
	During recent decades, a multitude of \ac{OSASI} algorithms has been developed which originated from the
	plain \acl{TD} \acl{LMS} algorithm \cite{widrow_b_adaptive_1960} and evolved to sophisticated implementations operating in the block-frequency domain \cite{ferrara_fast_1980, mansour_unconstrained_1982, haykin_2002}. In this context, robust \ac{OSASI} typically has to overcome interfering signals which are either undesired, e.g., \commentTHc{keyboard} noise, or desired, e.g., near-end talkers. Both types of interference are termed \textit{noise} in the following.
	
	%
	%
	%
	%
	Noisy observations are usually addressed by adaptation control mechanisms which are motivated by the non-stationarity of many acoustic excitation and noise signals. A popular approach for iterative \ac{OSASI} algorithms are \ac{VSS} control methods which lead to either binary or continuous-valued step sizes.
	As binary \ac{VSS} control, i.e., halting the filter adaptation during periods with high-level interfering noise \cite{Benesty_new_2000, haensler2004acoustic, cahill_approach_2008}, does not allow for permanent filter optimization, it is less suited for time-varying acoustic environments and applications with persistently high-level interfering noise signals.
	Thus, we focus in this paper on continuous \ac{VSS} control which stipulates a permanent filter adaptation.
	Many \acp{VSS} have been developed, ranging from scalar time-domain step sizes \cite{benesty-vss-lms, 6112248, 7115075}, to frequency-dependent step sizes which take into account the temporal correlation of the input and the noise signals \cite{nitsch2000frequency, enzner_frequency-domain_2006, malik_online_2010} and thus often result in faster convergence.
	%
	In particular, the inference of the adaptive filter coefficients by a Kalman filter \cite{7115075, enzner_frequency-domain_2006, malik_online_2010} has proven to be a powerful approach to \ac{VSS} control. The robustness of these algorithms against interfering signals crucially depends on precise estimates of the \commentTHa{process and observation noise power}, respectively. In \cite{malik_online_2010} it is proposed to estimate both jointly with the adaptive filter coefficients by optimizing a single \ac{ML} objective function. \commentTHc{However, due to the high-dimensional linear Gaussian \ac{DFT}-domain state space model \cite{enzner_frequency-domain_2006, malik_online_2010}, this approach greatly overestimates the noise power after an abrupt system change occurred. This results in a slow reconvergence which has been analyzed theoretically in  \cite{yang_frequency-domain_2017}.}

	In this paper we address this problem by introducing a nonnegative noise dictionary model which we term \ac{SSFDAF-NMF}. The proposed model captures the pronounced spectral structure which characterizes many types of {interfering} noise signals, e.g., wind noise \cite{schmidt_wind_2007}, music \cite{fevotte_nonnegative_2009}, speech \cite{smaragdis_convolutive_2007} or robotic ego-noise \cite{haubner_multichannel_2018}. 
	We suggest to estimate the noise dictionary from training data and infer continuously its activation by a generalized \ac{EM} algorithm \cite{dempster_maximum_1977}. 
	The optimization of the model shows a close relation to \ac{IS} divergence-based \acs{NMF}. 
	%
	%
	The proposed algorithm represents a computationally efficient block-online supervised adaptive filter with inherent noise-robust \ac{VSS} control \commentTHc{which allows for faster reconvergence after abrupt system changes.} 
	
	We use bold lowercase letters for vectors and bold uppercase letters for matrices with underlined symbols indicating \acl{TD} quantities. %
	%
	The $D$-dimensional identity matrix is denoted by $\boldsymbol{I}_D$, the $D$-dimensional \ac{DFT} matrix by $\boldsymbol{F}_D$ and the zero-matrix of dimensions $D_1 \times D_2$ by $\boldsymbol{0}_{D_1 \times D_2}$. 
	The superscripts $(\cdot)^{\text{T}}$ and $(\cdot)^{\text{H}}$ represent transposition and Hermitian transposition, respectively. We denote the determinant by $\text{det}(\cdot)$, the trace by $\text{Tr}(\cdot)$, and introduce the $\text{diag}(\cdot)$ operator which creates a diagonal matrix from its vector-valued argument. Finally, we use $\stackrel{\text{c}}{=}$ to denote equality up to a constant and $\circ$ for indicating element-wise operations.
	%
	
	\section{Online {Maximum-Likelihood} Learning}
	\label{sec:onlineML_learning}
	As the proposed algorithm is based on the noisy linear \acl{FD} echo path model described in \cite{enzner_frequency-domain_2006}, and related to the online inference of its parameters introduced in \cite{malik_online_2010}, we give in the following section a short summary. 
	
	\subsection{Linear Gaussian DFT-Domain State-Space Model}
	\label{sec:stateSpaceMod}
	%
	We model the noisy $R$-dimensional \acl{TD} observation vector
	%
	\begin{equation}
		{\underline{\boldsymbol{y}}}_\tau = \boldsymbol{Q}_1^{{\text{T}}} \boldsymbol{F}_M^{-1} \boldsymbol{X}_\tau \boldsymbol{F}_M \boldsymbol{Q}_2 \underline{\w}_\tau + \underline{\bs}_\tau \in \mathbb{R}^{R} 
		\label{eq:timeDomObsEq}
	\end{equation}
	at block-index $\tau$ by an overlap-save convolution of the input signal
	\begin{equation}
		\underline{\boldsymbol{x}}_\tau = \begin{pmatrix}
			\underline{x}_{\tau R - M + 1}, \underline{x}_{\tau R - M + 2}, \dots, \underline{x}_{\tau R}
		\end{pmatrix}^{{\text{T}}} \in \mathbb{R}^{M} 
		\label{eq:timeDomInSig}
	\end{equation}
	of even length $M$ and block-shift $R$ with the \ac{FIR} filter $\underline{\w}_\tau \in \mathbb{R}^{M-R}$ which is superimposed by the noise vector
	\begin{equation}
		\underline{\bs}_\tau = \begin{pmatrix}
			\underline{s}_{\tau R - R + 1}, \underline{s}_{\tau R - R + 2}, \dots, \underline{s}_{\tau R}
		\end{pmatrix}^{{\text{T}}} \in \mathbb{R}^{R}. 
	\end{equation}
	We used here the diagonal matrix \makebox{$\X_{\tau} = \text{diag} \left( \boldsymbol{F}_M \underline{\boldsymbol{x}}_{\tau} \right) \in \mathbb{C}^{M \times M}$}, which represents the \ac{DFT}-transformed input signal block $\underline{\boldsymbol{x}}_\tau$, the linear convolution constraint matrix \makebox{$\boldsymbol{Q}_1^{\text{T}} = \begin{pmatrix}\boldsymbol{0}_{R \times M-R} & \boldsymbol{I}_R\end{pmatrix}$} and the zero-padding matrix \makebox{$\boldsymbol{Q}_2^{\text{T}}= \begin{pmatrix}\boldsymbol{I}_{M-R} & \boldsymbol{0}_{M-R \times R }\end{pmatrix}$}.
	%
	%
	%
	By transforming the zero-padded \acl{TD} signals in Eq.~\eqref{eq:timeDomObsEq} to the \ac{DFT} domain, i.e.,
	\begin{equation}
		{\y}_\tau = \boldsymbol{F}_M \boldsymbol{Q}_1 {\underline{\boldsymbol{y}}}_{\tau} \in \mathbb{C}^{M} \quad\text{and}\quad {\bs}_\tau = \boldsymbol{F}_M \boldsymbol{Q}_1 \underline{\bs}_\tau \in \mathbb{C}^{M}, \label{eq:FD_obsDef}
	\end{equation}
	we obtain the {\acl{FD}} observation equation
	\begin{equation}
		{\y}_{\tau} = {\C}_{\tau}  {\w}_{\tau} + {\bs}_{\tau} 
	\end{equation}
	with the \ac{DFT}-transformed \ac{FIR} filter ${\w}_{\tau} = \boldsymbol{F}_M \boldsymbol{Q}_2 \underline{\w}_\tau \in \mathbb{C}^M$ and the overlap-save constrained input signal \makebox{$\boldsymbol{C}_\tau = \boldsymbol{F}_M \boldsymbol{Q}_1 \boldsymbol{Q}_1^{\text{T}} \boldsymbol{F}_M^{-1} \X_{\tau}$}.
	Here, we model the \acl{FD} observation noise vector ${\bs}_{\tau}$ as non-stationary, block-wise and spectrally uncorrelated zero-mean complex Gaussian random process which is distributed according to the \ac{pdf}
	\begin{equation}
		p({\bs}_\tau) = p({\bs}_\tau|{\boldsymbol{S}}_{1:\tau-1})  = \mathcal{N}_c({\bs}_{\tau}|\boldsymbol{0}_{M \times 1}, {\boldsymbol{\Psi}}_{\tau}^S), \label{eq:obsNoiseDistrib}
	\end{equation}
	with ${\boldsymbol{S}}_{1:\tau-1} = \begin{pmatrix}{\bs}_{1}, & \dots, & {\bs}_{\tau-1}\end{pmatrix}$. The diagonal entries of the noise covariance matrix $\left[\boldsymbol{\Psi}_{\tau}^S\right]_{mm} = \mathbb{E}\left[{s}_{m \tau} {s}_{m \tau}^* \right]$, with $\mathbb{E}\left[\cdot\right]$ denoting the expectation operator, approximate the noise \ac{PSD} at block $\tau$.
	%
	
	In \cite{enzner_frequency-domain_2006}, it is proposed to model the temporal evolution of the \ac{DFT}-transformed \ac{FIR} filter ${\w}_{\tau}$ in terms of a random-walk Markov model with a stationary and diagonal process noise covariance matrix $\boldsymbol{\Psi}^\Delta_{\tau}$. This allows to describe the \ac{OSASI} problem by the linear Gaussian \ac{DFT}-domain state space model 
	\begin{align}
		{\w}_{\tau} &= A  ~{\w}_{\tau - 1}  + \Delta {{\w}}_{\tau}  &\text{with}& \quad  \Delta {{\w}}_{\tau} \sim   \mathcal{N}_c(\Delta {{\w}}_{\tau}|{\boldsymbol{0}_{M \times 1}}, \boldsymbol{\Psi}_{\tau}^\Delta)\label{eq:stateTransMod} 
		\\
		{\y}_{\tau} &= {\C}_{\tau}  {\w}_{\tau} + {\bs}_{\tau} &\text{with} & ~~~\quad {\bs}_{\tau} \sim   \mathcal{N}_c({\bs}_{\tau}|{\boldsymbol{0}_{M \times 1}}, \boldsymbol{\Psi}_{\tau}^S), \label{eq:obsMod}
	\end{align}
	with the state transition coefficient \makebox{$0 < A < 1$}.

	\subsection{Online Inference}
	\label{sec:probInf}
	In \cite{malik_online_2010} it is proposed to infer the state-space model parameters in \makebox{$\tilde{\Theta}_\tau = \{\boldsymbol{\Psi}_{\tau}^S, \boldsymbol{\Psi}_{\tau}^\Delta \}$} by optimizing the \ac{ML} objective function
	%
	%
	\begin{align}
		\label{eq:mlMCSSFDAF}
		\begin{split}
			\tilde{\mathcal{C}}_{\text{ML}}(\tilde{\Theta}_\tau) = \log p({\y}_{\tau} | {\boldsymbol{Y}}_{1:\tau-1}, \tilde{\Theta}_\tau) 
		\end{split}
	\end{align}
	with ${\boldsymbol{Y}}_{1:\tau-1} = \begin{pmatrix}{\boldsymbol{y}}_{1}, & \dots, & {\boldsymbol{y}}_{\tau-1}\end{pmatrix}$ by an \ac{EM} algorithm which we term \ac{SSFDAF}. The adaptive filter coefficient vector ${\w}_{\tau}$ is treated as latent random vector of the log-likelihood \eqref{eq:mlMCSSFDAF} which leads via Jensen's inequality to the lower bound \cite{malik_online_2010} 
	%
	\begin{align}
		\tilde{\mathcal{C}}_{\text{ML}}(&\tilde{{\Theta}}_\tau) = \log \int p({\y}_{\tau}, {\w}_{\tau} | {\boldsymbol{Y}}_{1:\tau-1}, \tilde{\Theta}_\tau)~d {\w}_\tau \label{eq:mlMCSSFDAFNMF1}  \\
		&\geq  \int q({\w}_{\tau}) \log \frac{p({\y}_{\tau}, {\w}_{\tau} | {\boldsymbol{Y}}_{1:\tau-1}, \tilde{\Theta}_{\tau})}{q({\w}_{\tau})} ~d {\w}_\tau 
		= \tilde{\mathcal{Q}}(q,\tilde{\Theta}_\tau) \notag
	\end{align}
	for the log-likelihood function with $q({\w}_{\tau})$ being some \ac{pdf}.
	The $l$-th E-step of the algorithm, i.e., the variational optimization w.r.t. $q({\w}_{\tau})$, is addressed by the Kalman filter update 
	\vspace*{-.07cm}
	\begin{align} 
		\label{eq:eStep}
		\begin{split}
			\hat{\w}^{+}_{\tau - 1} &= A~\hat{\w}_{\tau - 1,(L)} \\
			{\bP}^{+}_{\tau - 1,(l)} &= A^2 ~ {\bP}_{\tau - 1,(L)} + \boldsymbol{\Psi}^{\Delta}_{\tau,(l)}\\
			\boldsymbol{\Lambda}_{\tau,(l)} &= {\bP}^{+}_{\tau - 1,(l)} \left( {\X}_\tau {\bP}^{+}_{\tau-1,(l)} {\X}_\tau^{\herm}  + \frac{M}{R} \boldsymbol{\Psi}^{S}_{\tau,(l)}
			\right)^{-1}\\
			\commentTHa{{\e}_{\tau}^{+}} &= \commentTHa{{\y}_\tau - {\C}_\tau \hat{\w}^{+}_{\tau - 1}} \\
			\hat{\w}_{\tau,(l)} & = \hat{\w}^{+}_{\tau - 1} + \boldsymbol{\Lambda}_{\tau,(l)} {\X}_\tau^{\herm} \commentTHa{{\e}_{\tau}^{+}} \\
			{\bP}_{\tau,(l)} & =  \left[{\I}_M - \frac{R}{M} \boldsymbol{\Lambda}_{\tau,(l)} {\X}_\tau^{\herm} {\X}_\tau\right] {\bP}^{+}_{\tau-1,(l)}
		\end{split}
	\end{align}
	%
	%
	\commentTHa{with the prior error ${\e}_{\tau}^{+}$, the posterior mean $\hat{\w}_{\tau,(l)}$ and the corresponding diagonal state uncertainty ${\bP}_{\tau,(l)}$ \cite{enzner_frequency-domain_2006}.}
	Note that the frequency-dependent step sizes, contained in the diagonal matrix $\boldsymbol{\Lambda}_{\tau,(l)}$, depend on the predicted state uncertainty ${\bP}^{+}_{\tau - 1,(l)}$, the \ac{DFT}-domain input signal ${\X}_\tau$ and on the estimated noise covariance matrix $\boldsymbol{\Psi}^{S}_{\tau,(l)}$. By inserting the optimum function $q^{(l)}$, i.e., the posterior \ac{pdf} resulting from the Kalman filter Eqs.~\eqref{eq:eStep}, into \eqref{eq:mlMCSSFDAFNMF1} we obtain the lower bound \cite{malik_online_2010}
	\begin{align}
		\tilde{\mathcal{Q}}( q^{(l)}, \tilde{\Theta}_\tau) \stackrel{\text{c}}{=} \tilde{\mathcal{Q}}_{\Delta}(q^{(l)},{\boldsymbol{\Psi}}_\tau^\Delta) + \tilde{\mathcal{Q}}_S(q^{(l)},{\boldsymbol{\Psi}}_\tau^S)
		\label{eq:addLowBound}
	\end{align}
	with  
	\begin{align}
		\tilde{\mathcal{Q}}_{\Delta}(&{q^{(l)},\boldsymbol{\Psi}}_\tau^\Delta)  \stackrel{\text{c}}{=}  - \log {\text{det} \left(  \boldsymbol{P}^+_{\tau-1} \right)}  \\  & -\text{Tr} \left(\left( \boldsymbol{P}_{\tau-1}^{+}\right)^{-1} \mathbb{E}_{q^{(l)}} \left[ \left( {\w}_\tau - A \hat{\w}_{\tau-1} \right) \left( {\w}_\tau - A \hat{\w}_{\tau-1} \right)^{\text{H}} \right] \right) \notag \\
		\tilde{\mathcal{Q}}_{S}(&{q^{(l)}, \boldsymbol{\Psi}_\tau^S})  \stackrel{\text{c}}{=} - \log {\text{det} \left(  \boldsymbol{\Psi}_\tau^S  \right)} \label{eq:eqlowBoundObsNoise}  \\
		&-\text{Tr} \left( {\left( \boldsymbol{\Psi}_\tau^S \right)}^{-1} \mathbb{E}_{q^{(l)}} \left[ \left( {\y}_\tau - {\C}_\tau {\w}_\tau\right) \left( {\y}_\tau - {\C}_\tau {\w}_\tau\right)^{\text{H}} \right] \right). \notag 
	\end{align}
	%
	Finally, in the M-step the optimum parameters in \makebox{$\tilde{\Theta}_\tau = \{\boldsymbol{\Psi}_{\tau}^S, \boldsymbol{\Psi}_{\tau}^\Delta \}$} are obtained by \cite{malik_online_2010}
	\begin{align}
		\boldsymbol{\Psi}_{\tau,(l)}^{\Delta} & = (1-A^2)~  \left(\hat{\w}_{\tau,(l)} {\hat{\w}_{\tau,(l)}}^{\text{H}} + {\bP}_{\tau,(l)}  \right)	\label{eq:mStepStateNoise}\\
		\boldsymbol{\Psi}_{\tau,(l)}^{S} & = {\e}_{\tau,(l)} {\e}_{\tau,(l)}^{\text{H}} + \frac{R}{M} {\X}_{\tau}  {\bP}_{\tau,(l)} {\X}_{\tau}^{\herm} \label{eq:mStepObsNoise}
	\end{align}
	with the \commentTHa{posterior} error ${\e}_{\tau,(l)} = {\y}_\tau - {\C}_{\tau} \hat{\w}_{\tau,(l)}$. 	
	\commentTHc{Another common choice for estimating the noise covariance matrix is given by}
	\begin{equation}
		\boldsymbol{\Psi}_{\tau,(l)}^{S} = \lambda \boldsymbol{\Psi}_{\tau-1,(L)}^{S} + (1-\lambda) \boldsymbol{\Psi}_{\tau,\text{prior}}^{S}  \label{eq:mStepObsNoiseRecAv}
	\end{equation}	
	\commentTHa{with the prior rank-1 error update $ \boldsymbol{\Psi}_{\tau,\text{prior}}^{S} = {\e}_{\tau}^{+} ({\e}_{\tau}^{+})^{\text{H}}$ \cite{franzen_improved_2019}. This can be motivated as a heuristic variant of \eqref{eq:mStepObsNoise} by omitting the excitation signal contribution and using the predicted mean $\hat{\w}^{+}_{\tau - 1}$ as initial estimate of the updated mean $\hat{\w}_{\tau,(l)}$.} By assuming $\boldsymbol{\Psi}_{\tau,(l)}^{\Delta}$ and $\boldsymbol{\Psi}_{\tau,(l)}^{S}$ to be diagonal, the off-diagonal terms of the outer products in \commentTHa{Eqs.~\eqref{eq:mStepStateNoise}, \eqref{eq:mStepObsNoise} and \eqref{eq:mStepObsNoiseRecAv}} are neglected.

	\section{Proposed \ac{SSFDAF-NMF} Algorithm}
	\label{sec:propAlg}
	In this section we describe the proposed \ac{VSS} control by exploiting a nonnegative dictionary noise model.
	

	\subsection{Probabilistic {Nonnegative Dictionary} Noise Model}
	\label{sec:probMod}
	Here, in contrast to \cite{malik_online_2010}, and all other state-of-the-art \ac{VSS} control strategies, we model the observation noise covariance matrix by a nonnegative dictionary model
	%
	%
	\begin{equation}
		\boldsymbol{\Psi}_\tau^S = \text{diag}\left({\dicMat} {\actVec}_{\tau} \right).
		\label{eq:nmfObsCovMod} 
	\end{equation}
	Hence, the observation model \eqref{eq:obsMod} is parametrized by the nonnegative dictionary matrix \makebox{${\dicMat} \in \mathbb{R}^{M \times K}_{\geq0}$} which contains $K$ noise atoms and the respective activation vector \makebox{$\actVec_\tau \in \mathbb{R}^{K}_{\geq0}$}. 
	We suggest to employ only the activation vector ${\actVec}_{\tau}$ of the \commentTHd{observation} noise model and the \commentTHd{process} noise covariance matrix {${\boldsymbol{\Psi}}_{\tau}^\Delta$} as state-space model parameters \makebox{${\Theta}_\tau = \{{\actVec}_{\tau}, {\boldsymbol{\Psi}}_{\tau}^\Delta \}$} which need to be inferred continuously online from the noisy observations. 
	The dictionary matrix ${\dicMat}$ is estimated from a {\acl{TD}} training data vector \makebox{$\underline{\bs}_{\text{tr}}^{\text{T}} = \begin{pmatrix} \underline{\bs}_{1}^{\text{T}} &, \dots& \underline{\bs}_{J}^{\text{T}}  \end{pmatrix} \in \mathbb{R}^{JR}$} which can be obtained either offline, if prior knowledge about the expected noise type is available, or online otherwise. In the latter case we append the observed signal $\underline{\boldsymbol{y}}_\tau$ to $\underline{\bs}_{\text{tr}}$, i.e., $\underline{\bs}_{\text{tr}}^{\text{T}} \gets \begin{pmatrix} \underline{\bs}_{\text{tr}}^{\text{T}}, \underline{\boldsymbol{y}}_\tau^{\text{T}}  \end{pmatrix}$, whenever the input signal is not active, i.e., $\underline{\boldsymbol{x}}_\tau\approx \boldsymbol{0}_{M \times 1}$, and thus $\underline{\boldsymbol{y}}_\tau \approx \underline{\bs}_\tau$ (cf.~Eq.~\eqref{eq:timeDomObsEq}).
	The \acl{TD} training data vector $\underline{\bs}_{\text{tr}}$ is straightforwardly transformed to the corresponding \ac{STFT} training data matrix \makebox{${\boldsymbol{S}}_{\text{tr}} = \begin{pmatrix}{\bs}_{\text{tr},1} & \dots & {\bs}_{\text{tr},N} \end{pmatrix} \in \mathbb{C}^{M \times N}$}, by decomposing $\underline{\bs}_{\text{tr}}$ into $N$ blocks $\underline{\bs}_{\text{tr},\tau} \in \mathbb{R}^M$ of length $M$ with block shift $R_{\text{tr}}$, followed by windowing and \ac{DFT} transformation. Note that in comparison to $\underline{\bs}_\tau$ (cf.~Eq.~\eqref{eq:FD_obsDef}), the time-domain training data blocks $\underline{\bs}_{\text{tr},\tau}$ are not zero-padded to increase the spectral resolution of the corresponding non-stationary noise \ac{PSD} samples $|{\bs}_{\text{tr},\tau}|^{\circ 2}$ where $\tau=1,\dots,N$.
	%
	%
	%
	We propose to estimate the dictionary matrix $\dicMat$ by maximizing the log-likelihood $\log p(\boldsymbol{S}_{\text{tr}}) = \prod_{\tau=1}^{N} p({\bs}_{\text{tr},\tau})$ which is equivalent to maximizing the \acs{IS-NMF} objective function \cite{fevotte_nonnegative_2009}
	%
	\begin{equation}
		\mathcal{C}_{\text{IS}}({\dicMat}, \boldsymbol{V}_{\text{tr}})  \stackrel{\text{c}}{=}   - \hspace*{-.1cm} \sum_{m,\tau=1}^{M,N} \left(\log  \sum_{k=1}^{K} t_{m k} v_{\text{tr},k \tau} + \frac{ |{s}_{\text{tr},m\tau}|^2 }{\sum_{k=1}^{K} t_{m k} v_{\text{tr},k \tau}} \right) \label{eq:isNMFcostFunc} 
	\end{equation}
	with $s_{\text{tr},m \tau} = \left[ \boldsymbol{S}_{\text{tr}} \right]_{m \tau}$, $t_{mk} = \left[ \dicMat \right]_{mk} $ and $v_{\text{tr},k \tau} = \left[ \boldsymbol{V}_{\text{tr}} \right]_{k \tau}$.
	%
	This allows to optimize the dictionary matrix ${\dicMat}$ by the multiplicative \acs{IS-NMF} update rules \cite{fevotte_algorithms_2011}
	\begin{align}
		{\boldsymbol{V}_{\text{tr}}} &\gets {\boldsymbol{V}_{\text{tr}}} \circ \left[ \frac{{\dicMat}^{\text{T}} \left( ({\dicMat} {\boldsymbol{V}_{\text{tr}}})^{\circ -2} \circ |{\boldsymbol{S}}_{\text{tr}}|^{\circ 2} \right)}{{\dicMat}^{\text{T}} ({\dicMat} {\boldsymbol{V}_{\text{tr}}})^{\circ -1}  }  \right]^{\circ\frac{1}{2}} \label{eq:mm-update-act} \\
		{\dicMat} &\gets {\dicMat} \circ \left[ \frac{ \left( ({\dicMat} {\boldsymbol{V}_{\text{tr}}})^{\circ -2} \circ |\boldsymbol{S}_{\text{tr}}|^{\circ 2} \right)  {\boldsymbol{V}_{\text{tr}}}^{\text{T}}}{ ({\dicMat} {\boldsymbol{V}_{\text{tr}}})^{\circ -1} {\boldsymbol{V}_{\text{tr}}}^{\text{T}}  }  \right]^{\circ\frac{1}{2}} \label{eq:mm-update-dic}
		%
	\end{align}
	which represent an instance of the \ac{MM} algorithm \cite{hunter2004tutorial}. Note that the training data activation matrix $\boldsymbol{V}_{\text{tr}}$ is only needed for estimating the dictionary and that due to the conjugate symmetry of ${\bs}_\tau$ and ${\bs}_{\text{tr},\tau}$, it is sufficient to {compute} the non-redundant part of the dictionary, i.e., \makebox{$\tilde{\dicMat} = \begin{pmatrix}\boldsymbol{I}_{\frac{M}{2}+1} & \boldsymbol{0}_{\frac{M}{2}+1 \times \frac{M}{2}-1}\end{pmatrix} {\dicMat}$}. 

	\subsection{Online Inference}
	\label{sec:inference2}
	%
	We propose to estimate the parameters in ${\Theta}_\tau$ by maximizing the log-likelihood function
	\begin{align}
		\label{eq:mlMCSSFDAFNMF}
		\begin{split}
			\mathcal{C}_{\text{ML}}({\Theta}_\tau) &= \log p({\y}_{\tau} | {\boldsymbol{Y}}_{1:\tau-1}, {\Theta}_\tau).
		\end{split}
	\end{align}
	Due to the linear model, we obtain a tight lower bound on the activation vector $\actVec_{\tau}$ by inserting the dictionary model \eqref{eq:nmfObsCovMod} into \eqref{eq:eqlowBoundObsNoise} and evaluating the expectation
	%
	%
	\begin{equation}
		\mathcal{Q}_{S}(q^{(l)},{\actVec}_\tau)  \stackrel{\text{c}}{=} - \sum_{m=1}^{M} \left(\log  \sum_{k=1}^{K} t_{m k} v_{k \tau} + \frac{\psi_{m \tau,(l)}^e }{\sum_{k=1}^{K} t_{m k} v_{k \tau}} \right) \label{eq:boundForNoiseSimple} 
	\end{equation}
	with $\psi_{m \tau,(l)}^e = \left[ {\e}_{\tau,(l)} {\e}_{\tau,(l)}^{\text{H}} + {\C}_\tau \boldsymbol{P}_{\tau,(l)} {\C}_\tau^{\text{H}}  \right]_{mm}$ representing the expected posterior error power. By comparing Eq.~\eqref{eq:boundForNoiseSimple} to Eq.~\eqref{eq:isNMFcostFunc}, we observe that the lower bound on the activation vector ${\actVec}_\tau$ is an \acs{IS-NMF} objective function with the expected posterior error power $\psi_{m \tau,(l)}^e$ as target variable. 
	%
	The lower bound $\mathcal{Q}_{S}(q^{(l)},{\actVec}_\tau)$ does, in comparison to $\tilde{\mathcal{Q}}_{S}(q^{(l)},{\boldsymbol{\Psi}}_\tau^S)$ ({cf.}~Eq.~\eqref{eq:eqlowBoundObsNoise}), not allow for a closed-form solution as given by~Eq.~\eqref{eq:mStepObsNoise}. Thus, we suggest to optimize it iteratively by applying $P$ times the multiplicative update \eqref{eq:mm-update-act}\commentTHd{, with ${\boldsymbol{V}_{\text{tr}}} = {\actVec}_\tau$ and $|{s}_{\text{tr},m\tau}|^2=\psi_{m \tau,(l)}^e$,} which yields a series of non-decreasing values of the objective function \cite{fevotte_algorithms_2011}. 
	%
	Hence, the proposed \ac{SSFDAF-NMF} represents a generalized \ac{EM} algorithm {\cite{dempster_maximum_1977}}. 

	For updating the process noise covariance matrix ${\boldsymbol{\Psi}}_\tau^\Delta$, we first observe that the optimization w.r.t. ${\boldsymbol{\Psi}}_\tau^\Delta$ is independent of ${\actVec}_\tau$ which results from the additive structure of the lower bound \eqref{eq:addLowBound}. Hence, we can use Eq.~\eqref{eq:mStepStateNoise}.

	\subsection{\commentTHa{Algorithmic Variants}}
	\label{sec:algDescr}
	\commentTHa{For updating the filter coefficients, two different optimization procedures are investigated}. Either the E-step \eqref{eq:eStep} \commentTHa{is carried out} before the M-step \eqref{eq:mStepStateNoise} and \eqref{eq:mStepObsNoise}, \commentTHa{abbreviated by EM, or the reversed order \commentTHa{is used}, abbreviated by ME.} 
	%
	%
	\commentTHa{Note that the EM version, which is termed SSFDAF-EM-$L$ in the following, requires at least $L=2$ iterations to use the updated noise covariance matrix $\boldsymbol{\Psi}_{\tau,(l)}^{S}$ (cf.~Eq.~\eqref{eq:mStepObsNoise}) in the Kalman step size computation (cf.~Eq.~\eqref{eq:eStep}).} 
	
	\commentTHa{The proposed \ac{SSFDAF-NMF} update with the EM optimization version including $L$ iterations is summarized in Alg.~\ref{alg:prop_alg_descr}.} 
	\commentTHa{Here, for each E-step the posterior mean $\hat{\w}_{\tau,(l)}$ and the respective state uncertainty matrix $\boldsymbol{P}_{\tau,(l)}$ of the latent adaptive filter coefficients are updated by the Kalman filter Eqs.~\eqref{eq:eStep}. \commentTHd{Subsequently, in the M-step, the diagonal process noise covariance matrix $\boldsymbol{\Psi}_{\tau,(l)}^{\Delta}$ and dictionary activation ${\actVec}_{\tau,(l)}$ are updated. The dictionary activation is optimized by applying $P$ times the multiplicative update rule \eqref{eq:mm-update-act} with the target variable \makebox{$\psi_{m \tau,(l)}^e$} (cf.~Sec.~\ref{sec:inference2}) and fixed dictionary matrix $\dicMat$. Note that the expected posterior error power \makebox{$\psi_{m \tau,(l)}^e$} can efficiently be approximated by Eq.~\eqref{eq:mStepObsNoise} \makebox{$\psi_{m \tau,(l)}^e \approx  \left[ \boldsymbol{\Psi}_{\tau,(l)}^{S}  \right]_{mm}$} \cite{enzner_frequency-domain_2006}.}} 
	%
	%
	\commentTHc{The iterative optimization procedure is initialized with the optimum activation vector $\boldsymbol{v}_{\tau-1,(L)}$ of the previous time frame. Finally, the observation noise covariance matrix $\boldsymbol{\Psi}_{\tau,(l)}^S$ is updated by Eq. \eqref{eq:nmfObsCovMod}.}
	\begin{algorithm}[b] 
		\caption{\commentTHa{Proposed filter update by SSFDAF-NMF-EM-$L$.}} 
		%
		\label{alg:prop_alg_descr}
		\begin{algorithmic}
			\For{$l=1,\dots,L$}
			\State \hspace*{-.1cm}E-Step: Update $\hat{\w}_{\tau,(l)}$ and $\boldsymbol{P}_{\tau,(l)}$ by Kalman filt.~\eqref{eq:eStep}
			\State \hspace*{-.1cm}M-Step:
			\State \hspace*{.45cm}Update \commentTHd{process noise cov. matrix} $\boldsymbol{\Psi}_{\tau,(l)}^{\Delta}$ by Eq.~\eqref{eq:mStepStateNoise}
			\State \hspace*{.45cm}Compute exp. post. error power $\boldsymbol{\Psi}_{\tau,(l)}^{S}$ by Eq.~\eqref{eq:mStepObsNoise}
			\State \hspace*{.45cm}Optimize ${\actVec}_{\tau,(l)}$ by \commentTHa{repeating} Eq.~\eqref{eq:mm-update-act} $P$ times
			\State \hspace*{.45cm}Update \commentTHd{obs. noise cov. matrix} $\boldsymbol{\Psi}_{\tau,(l)}^S \gets \text{diag}\left({\dicMat} {\actVec}_{\tau,(l)}\right)$
			\EndFor	
		\end{algorithmic}
	\end{algorithm}
	
	\commentTHd{On the other hand,} \commentTHa{if using the ME version with $L=1$ iteration and the observation noise covariance estimator \eqref{eq:mStepObsNoiseRecAv}, one obtains the optimization approach described in \cite{franzen_improved_2019} which is termed SSFDAF-ME-$1$ in the sequel.}
	\commentTHa{We propose \commentTHd{here} an alternative NMF-based M-step by using the instantaneous prior error power, i.e., $\lambda=0$, as desired target variable of the NMF model \makebox{${\psi}_{m \tau,(l)}^e =\left[{\e}_{\tau}^{+} ({\e}_{\tau}^{+})^{\text{H}}\right]_{mm}$} (cf.~Eq.~\eqref{eq:boundForNoiseSimple}), which is abbreviated by SSFDAF-NMF-ME-$1$.}

	\section{Experiments}
	\label{sec:experiments}
	In this section the proposed \ac{SSFDAF-NMF} \commentTHc{variants are} evaluated for an \acl{AEC} scenario \commentTHc{which \commentTHa{includes} an abrupt echo path change and various types of recorded noise signals.} 
	The respective echoes \commentTHa{$\underline{\boldsymbol{y}}_{\text{ec},\tau}$} were simulated by convolving speech source signals, i.e., $15$ randomly-selected talkers of \cite{uwnu_corpus}, with measured \acp{RIR} \commentTHf{$\underline{\boldsymbol{h}}_{\tau}\in \mathbb{R}^D$} of length $D=48000$ and sampling frequency \makebox{$f_{\text{s}}=16$ kHz}. The \acp{RIR} were taken from the AIR database \cite{jeub_we_2010} and describe a hands-free interaction of a human with a mock-up phone in $8$ different acoustic environments. The echo path change is modelled \commentTHf{by using different \acp{RIR} $\underline{\boldsymbol{h}}_\tau$ for simulating the echoes before and after $10$ s}.
	%
	The clean echo signals \commentTHa{$\underline{\boldsymbol{y}}_{\text{ec},\tau}$} were distorted by adding recorded noise signals, which were scaled according to the desired \ac{SNR}. As noise signals we considered near-end speech, chosen from additional $15$ different talkers of \cite{uwnu_corpus}, \commentTHa{keyboard clicking and moving cutlery \cite{freesoundorg}}.
	%
	%
	
	\commentTHa{As performance measures we \commentTHf{use the signal-dependent} \ac{ERLE} ${\Gamma}_\tau$ and the system mismatch $\Upsilon_\tau$ \cite{enzner_acoustic_2014}}
	%
	\begin{align*}
		{\Gamma}_\tau = 10 \log_{10} \frac{\mathbb{E}\left[||\underline{\boldsymbol{y}}_{\text{ec},\tau}||_2^2\right] }{\mathbb{E}\left[||\underline{\boldsymbol{y}}_{\text{ec},\tau}-\hat{\boldsymbol{\underline{y}}}_{\tau}||_2^2\right]}, ~~
		\Upsilon_\tau = 10 \log_{10} \frac{{||\commentTHf{\underline{\boldsymbol{h}}_\tau} - \hat{\underline{\boldsymbol{w}}}_\tau}||_2^2}{||\commentTHf{{\underline{\boldsymbol{h}}}_{\tau}}||_2^2} \label{eq:systMisDef}
	\end{align*}
	%
	%
	\commentTHa{with $||\cdot||_2$ denoting the p2-norm, $\hat{\boldsymbol{\underline{y}}}_{\tau}$ denoting the estimated echo signal and $\hat{\underline{\boldsymbol{w}}}_\tau = \boldsymbol{Q}_2^{\text{T}} \boldsymbol{F}_M^{-1} \hat{{\boldsymbol{w}}}_\tau$.}
	%
	Note that we only use the first $M-R$ taps of the true \ac{RIR} \commentTHf{$\underline{\boldsymbol{h}}_\tau$} \commentTHa{for computing $\Upsilon_\tau$} to obtain an estimate of the attainable system mismatch. The echo caused by the remaining $D-(M-R)$ taps is usually dealt with by a residual echo suppression \commentTHa{filter}, e.g., \cite{haensler2004acoustic,enzner_frequency-domain_2006}, which we do not include here. 
	\commentTHc{As baseline approaches we use the state-of-the-art noise power estimators \eqref{eq:mStepObsNoise} \cite{malik_online_2010}, with EM optimization sequence and $L=2$ iterations (SSFDAF-EM-2), and \eqref{eq:mStepObsNoiseRecAv} \cite{franzen_improved_2019}, with $\lambda=0.5$ and $L=1$ iteration (SSFDAF-ME-1). The baseline algorithms are compared with two optimization variants of the proposed \ac{SSFDAF-NMF} algorithm with $K=10$ basis vectors and $P=3$ \ac{MM} optimization steps. While the first variant is defined by an an EM optimization sequence with $L=2$ iterations (SSFDAF-NMF-EM-2), the second one uses a single ME iteration and the prior error power approximation (SSFDAF-NMF-ME-1). For all algorithms we chose state transition coefficient $A=0.9999$, block-length $M=1536$ and block-shift $R=512$, corresponding to a filter length $M-R=1024$. } 
	%
	For training the dictionaries we chose a set of $25$ s-long signals which are temporally complementary to the evaluation data.
	%
	%
	%
	%
	%
	The near-end speech training data did neither include the input nor the local speech signal used for testing the algorithms \commentTHc{and thus allows for an offline speaker-independent training.}
	\begin{figure}[t!]
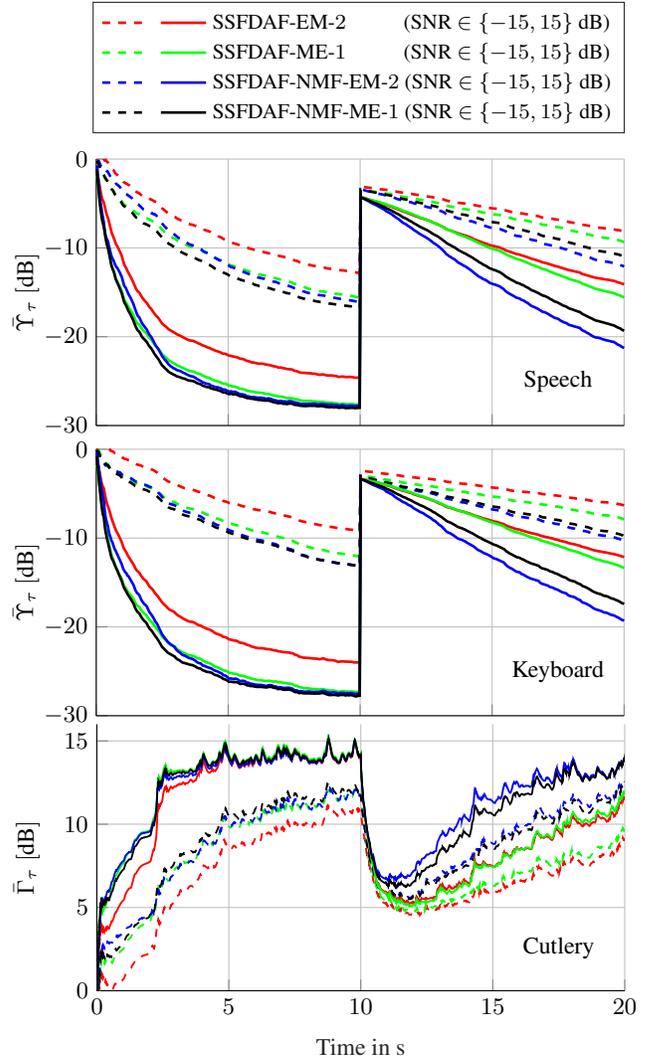

		\centering
		\vspace*{-.0cm}
		\hspace*{.47cm}
		\begin{subfigure}[t]{\columnwidth}
			\centering
			\newlength\fwidth
			\setlength\fwidth{1\columnwidth}
			\definecolor{mycolor2}{rgb}{0.00000,0.44700,0.74100}%
\definecolor{mycolor3}{rgb}{0.85000,0.32500,0.09800}%
\definecolor{mycolor1}{rgb}{0.92900,0.69400,0.12500}%

\begin{tikzpicture}
		\begin{axis}[%
		width=0.0\fwidth,
		height=0.0\fwidth,
		at={(0\fwidth,0\fwidth)},
		scale only axis,
		xmin=-20,
		xmax=10,
		xlabel style={font=\color{white!15!black}},
		ymin=-10,
		ymax=20,
		axis background/.style={fill=white},
		axis x line*=bottom,
		axis y line*=left,
		xmajorgrids,
		ymajorgrids,
		legend style={at={(0.3,0.0)}, anchor=south, legend cell align=left, align=left, draw=white!15!black, legend columns=2, font=\footnotesize}
		]

\addplot [color=red, dashed, line width=1.0pt]
table[row sep=crcr]{%
	-20	-9.71223617215463\\
};
\addlegendentry{}

\addplot [color=red, line width=1.0pt]
table[row sep=crcr]{%
	-20	-9.71223617215463\\
};
\addlegendentry{SSFDAF-EM-$2$ \hphantom{abcdd}\hspace*{.02cm}(SNR $\in\{-15,15\}$ dB)}

\addplot [color=green, dashed, line width=1.0pt]
table[row sep=crcr]{%
	-20	-9.71223617215463\\
};
\addlegendentry{}

\addplot [color=green, line width=1.0pt]
table[row sep=crcr]{%
	-20	-9.71223617215463\\
};
\addlegendentry{SSFDAF-ME-$1$ \hphantom{abcdd}\hspace*{.02cm}(SNR $\in\{-15,15\}$ dB)}	

\addplot [color=blue, dashed, line width=1.0pt]
table[row sep=crcr]{%
	-20	-9.71223617215463\\
};
\addlegendentry{}

\addplot [color=blue, line width=1.0pt]
table[row sep=crcr]{%
	-20	-9.71223617215463\\
};
\addlegendentry{SSFDAF-NMF-EM-$2$ (SNR $\in\{-15,15\}$ dB)}	

\addplot [color=black, dashed, line width=1.0pt]
table[row sep=crcr]{%
	-20	-9.71223617215463\\
};
\addlegendentry{}

\addplot [color=black, line width=1.0pt]
table[row sep=crcr]{%
	-20	-9.71223617215463\\
};
\addlegendentry{SSFDAF-NMF-ME-$1$ (SNR $\in\{-15,15\}$ dB)}	

		\end{axis}
		\end{tikzpicture}%
		\end{subfigure}
		
		\vspace*{.1cm}
		
		\begin{subfigure}[t]{\columnwidth}
			\centering
			\setlength\fwidth{.85\columnwidth}
			\input{systMis_exc-speech_noise-speech-speakIndepTrain_snr_15-room_2_3_4_5_6_7_8.tikz}
		\end{subfigure}
		
		\vspace*{-.15cm}
		
		\begin{subfigure}[t]{\columnwidth}
			\centering
			\setlength\fwidth{.85\columnwidth}
			\input{systMis_exc-speech_noise-keyboardmouse_snr_15-room_2_3_4_5_6_7_8.tikz}
		\end{subfigure}
		
		\vspace*{-.15cm}
		
		\hspace*{.03cm}
		\begin{subfigure}[t]{\columnwidth}
			\centering
			\setlength\fwidth{.85\columnwidth}
			\input{erle_exc-speech_noise-moving_cutlery_snr_15-room_2_3_4_5_6_7_8_save_save.tikz}
		\end{subfigure}
		\vspace*{-.5cm}
		\caption{\commentTHc{Performance evaluation of the proposed \ac{SSFDAF-NMF} algorithms in comparison to their baseline \ac{SSFDAF} counterparts for different noise types and \ac{SNR} levels.}}
		%
		\label{fig:resResults}
		
		\vspace*{-.0cm}	
	\end{figure}
	%
	%
	%
	%
	For computing the training data matrix $\boldsymbol{S}_{\text{tr}}$ we used a Hamming window and a frame-shift $R_{\text{tr}}=512$.

	\commentTHf{Fig.\ref{fig:resResults} shows exemplary average \ac{ERLE} $\bar{\Gamma}_\tau$ and system mismatch $\bar{\Upsilon}_\tau$ for different noise signal types and \ac{SNR} levels. Note that due to space constraints not all combinations can be shown.} 
	The results have been computed by evaluating {$100$} experiments with each experiment being defined by randomly drawing an \ac{RIR} \commentTHc{sequence}, an input and interfering noise signal and a training noise signal. 
	\commentTHc{As can be concluded from Fig.~\ref{fig:resResults}, the proposed \ac{SSFDAF-NMF} variants both \commentTHa{significantly} outperform their baseline counterparts in terms of reconvergence rate after echo path changes in all considered scenarios. Thus, the proposed noise dictionary allows the adaptive filter to \commentTHd{better} distinguish \commentTHa{whether} the estimated prior error $\boldsymbol{e}^+_\tau$ results from a noisy observation or \commentTHa{from} adaptive filter misalignment. Furthermore, the proposed {SSFDAF-NMF-EM-2} shows a better reconvergence behaviour than the proposed {SSFDAF-NMF-ME-1}, at the cost of slightly slower initial convergence rate.} 
	%
	While the benchmark algorithms (SSFDAF-EM-2) and (SSFDAF-ME-1) require on average $0.3$ ms and $0.6$ ms, respectively, to process one signal block $\boldsymbol{y}_\tau$ \commentTHa{of duration $32$ ms} on an \textit{Intel Xeon CPU E3-1275 v6 @ 3.80GHz}, their NMF counterparts need $0.5$ ms and $1$ ms, respectively. \commentTHf{As the runtime is still comparably low}, the proposed \ac{SSFDAF-NMF} algorithm represents a computationally highly efficient \ac{OSASI} algorithm.

	\section{Conclusion}
	\label{sec:summaryOutlook}
	We have introduced a novel computationally efficient adaptation control for block-\ac{OSASI} by exploiting a nonnegative noise dictionary. The proposed algorithm exhibits significantly faster convergence properties for different types of recorded noise signals in comparison to \commentTHf{two high-performance state-of-the-art algorithms.}

	\clearpage
	\vfill\pagebreak
	
	\bibliographystyle{IEEEbib}
	\bibliography{refs}

\end{document}